\documentclass{aastex}
\usepackage{graphics}
\usepackage{emulateapj5}
\usepackage{epsf}
\usepackage{color}

\def\lesssim{\mathrel{\hbox{\rlap{\hbox{\lower4pt\hbox{$\sim$}}}\hbox{$<$}}}}
\def\gtrsim{\mathrel{\hbox{\rlap{\hbox{\lower4pt\hbox{$\sim$}}}\hbox{$>$}}}}

\begin{document}

\title{The X-ray-Resolved Supernova Remnant S8 in the Dwarf Irregular
Galaxy IC 1613}

\author{E. M. Schlegel\altaffilmark{1,2}, Thomas
  G. Pannuti\altaffilmark{3}, T. Lozinskaya\altaffilmark{4},
  A. Moiseev\altaffilmark{5,6,7}, \& C. K. Lacey\altaffilmark{8}}

\altaffiltext{1}{Department of Physics and Astronomy, University of
Texas-San Antonio, San Antonio, TX 78249; eric.schlegel@utsa.edu}
\altaffiltext{2}{Vaughan Family Professor}
\altaffiltext{3}{Space Science Center, Department of Physics, Earth 
Sciences, and Space Systems Engineering, Morehead State University, Morehead, KY 40351;
t.pannuti@moreheadstate.edu}
\altaffiltext{4}{Sternberg Astronomical Institute, Moscow, 119992 Russia; lozinsk36@mail.ru}
\altaffiltext{5}{Special Astrophysical Observatory, Russian Academy of Sciences; moisav@sao.ru}
\altaffiltext{6}{Sternberg Astronomical Institute, Moscow, 119992 Russia}
\altaffiltext{7}{Space Research Institute, Russian Academy of Sciences, Profsoyuznaya ul. 84/32, Moscow 117997, Russia}
\altaffiltext{8}{Hofstra University, NY; christina.lacey@hofstra.edu}


\begin{abstract}

We conducted an observation of the nearby irregular galaxy IC 1613
with the {\it Chandra X-ray Observatory} using the S3 chip of the ACIS
with an effective exposure time of 49.9 ksec.  The observation
primarily targeted the extensive bubble and star formation region in
the northeast quadrant of the galaxy.  The only known supernova
remnant (SNR) in IC 1613, S8, is also the galaxy's most luminous X-ray
source (L$_{\rm X}$(0.3-8 keV) ${\sim}5-6{\times}10^{36}$ erg
s$^{-1}$).  We {\it resolve} the SNR with {\it Chandra} and compare
its nearly circular X-ray morphology with H${\alpha}$ and radio
images.  We assign an upper limit on the luminosity of any possible
associated compact central object of ${\approx}4{\times}10^{35}$ erg
s$^{-1}$ (0.3-8 keV band) and conclude that we would detect a
Crab-like pulsar but not a Cas A-like object.  We infer an age for S8
of ${\sim}$3400-5600 years and compare it to other SNRs in the Local
Group.  We suggest that S8 is a young composite SNR.

\end{abstract}


\section{Introduction}

Nearby galaxies are natural subjects of deep {\it Chandra}
observations because of {\it Chandra}'s high angular resolution
(90\% encircled energy within 1$''$).  {\it
Chandra} observations have dramatically increased the numbers of known
discrete X-ray sources in nearby galaxies (e.g., \cite{Long2010} for
M33; \cite{Holt2003} for NGC 6946, among many others).  The high accuracy of the
positions of detected X-ray sources permits searches for counterparts
at other wavelengths.

Supernova remnants (SNRs) represent one end stage for stellar
evolution of massive stars undergoing core collapse or white dwarfs
pushed past the Chandrasekhar mass limit.  Through a broad display of
explosion physics, SNRs potentially provide insight into the explosion
mechanism(s).  \cite{Vink2012} provides a recent review of X-ray
emission from MW SNRs.

Extragalactic SNRs are increasingly playing a role in broadening the
range of parameters beyond the MW examples (e.g., \cite{Long2010},
\cite{Pannuti2007}, \cite{Lacey1997}, \cite{MF1997}).  SNRs are
difficult to study in the Milky Way (MW) because their distances are
often uncertain and the intervening column of material can obscure
portions of their spectra.  The study of extragalactic SNRs is
particularly important for those host galaxies that are not spirals,
or for which the metallicity is very different, or the star formation
rate differs, among additional parameters.  Furthermore, distances are
a much smaller problem because the distance to the host galaxy is
often known to a high degree.  \cite{Leonidaki2017} provides a recent
review of this developing area.

In this paper, we describe a {\it Chandra} observation that {\it
resolves} an SNR in the nearby irregular Local Group galaxy IC 1613.
\cite{SchlegelIC1613} describes the overall results from the IC~1613
observation; this paper was extracted separately to attention to the
SNR.
The SNR, first noted by \cite{Sandage71} as an H~II region, was listed
in his paper as `S8' -- we adopt that label throughout this paper.
Additional details about S8 are described in \S~\ref{sect_snr}.

IC~1613 is classified as a DDO Irr~V \citep{vandenBergh00} lying
within the Local Group.  Estimates of its distance land around
$\sim$730 kpc (\citealt{Dolphin01}, \citealt{Karachentsev04}).  The
most recent distance determination is 724${\pm}$17 kpc \citep{Hatt17}.
We here adopt 725 kpc.  IC~1613 is known to be a low-metallicity
  galaxy ([Fe/H] = -1.3, \citealt{Kunth2000}).  The known Galactic
column density, N$_H$, in the direction of IC~1613 is
${\sim}2{\times}10^{20}$ cm$^{-2}$ based on the observed $E_{B-V}
{\sim}0.025$ \citep{Schlafly2011} and the low-metallicity relation of
\cite{Fitzpatrick1985}.  Given that the galaxy is a dwarf, its massive
star formation history is of interest.  This is particularly important
for IC~1613 given the very massive OB association in the northeast
quadrant of the galaxy and the relative lack of star clusters
elsewhere \citep{Wyder00}.

Prior observations of S8 are briefly reviewed in \S\ref{SNRpast}.  In
the X-ray band, \cite{Lozinskaya98} used the {\it ROSAT} HRI to
observe IC~1613 twice for $\sim$20 ks duration each.  Four sources
were observed: a background galaxy cluster, the SNR S8, an m$_v$ 11.4
foreground star, and a probable X-ray binary.  Sources 1, 3, and 4 of
that list are described in \cite{SchlegelIC1613}; here we describe the
second source.

\section{Observations and Data Reduction}\label{ObsSection}

We used the back-illuminated S3 CCD of {\it Chandra}'s Advanced
Charge-Coupled Device (CCD) Imaging Spectrometer (ACIS)
\citep{Gar2003} to observe IC 1613 on 2005 September 4 (obsid 5905)
for an effective exposure of 49327 sec.  The {\it Chandra} observation
covers roughly the central 1/2 of the galaxy.  The star
formation/bubble complex in the northeast quadrant of IC 1613 was the
target of the observation.  The SNR lies south of the center of the
complex and within the region of the CCD with the sharpest focus.

The Chandra Interactive Analysis of Observations (CIAO) software
(version 4.9) and the associated calibration files (version 4.7) were
used in the analysis.  To test for the presence of soft background
flares, we accumulated a source-free background area offset south from
the galaxy center and away from the galaxy cluster.  We extracted a
light curve using 50-second bins; no flares were detected.

The SNR events were extracted in two ways -- to obtain a spectrum and
to obtain its color-color information.  For the spectrum, we extracted
the counts in the 0.3-8 keV band from a region surrounding the SNR
with a radius of 4.5 arc seconds and centered at RA 01:05:02.4, Dec
+02:08:42.1.  A source-free region from the same CCD and $0'.5$ south
of the SNR was extracted for the background.  Response matrices and
effective area files were separately generated for the SNR and the
background.

Additional details of the observation and the data reduction are
presented in \cite{SchlegelIC1613}.  Those details are mostly related
to the other sources detected in IC~1613 and are not useful for
discussions of S8.

\section{The SNR}\label{sect_snr}

\subsection{Overview of Past Observations}\label{SNRpast}

First identified and cataloged by \cite{Sandage71}, S8 lies in the
northeast quadrant of IC~1613.  \cite{Smith75} noted strong [S~II]
lines in the spectrum of S8, a characteristic of SNRs and not H~II
regions.  Subsequently, \cite{DOdorico80} and \cite{Peimbert88}
confirmed the object as an SNR based on the high [S~II]/H${\alpha}$
ratio.  Radio emission from S8 was first detected by \cite{Dickel85}
from which those authors derived a spectral index of ${\alpha} =
-0.9\pm0.3$, a value that is associated with a synchrotron origin,
which is the characteristic emission process at radio wavelengths of
SNRs.  \cite{Lozinskaya98}, synthesized the available X-ray, optical,
and radio data and suggested S8 exploded in a cavity within a dense
H~I shell -- the galaxy is very rich in H~I with $\gtrsim$20\% of its
mass \citep{Huchtmeier81}.  They also measured a spectral index
${\alpha} = 0.57{\pm}0.054$; that value is consistent with
\cite{Dickel85}.

The overall reddening of the galaxy is E$_{B-V} {\sim} 0.04$
\citep{Freedman1988}.  \cite{Skillman2014} also note the low
foreground and internal reddening and \cite{Pietrzynski06} infer that
the reddening law is similar to that of the Milky Way.  In the
northeast complex of H~II regions where S8 is located, however, a
gradient exists in the $H{\alpha}/H{\beta}$ ratio.
\cite{Lozinskaya98} argued that a reddening of 0.2-0.4 in
C(H${\beta}$) produced the best-fit value, yielding an $E_{B-V}$ of
${\sim}0.12-0.24$ and an $N_H {\sim}1-2{\times}10^{22}$ cm$^{-2}$
based on the $N_H-E_{B-V}$ relation of \cite{Fitzpatrick1985}.

Differing estimates of the reddening directly impact the inferred
optical and X-ray luminosity and hence estimates of the electron
density, $n_e$.  Based on the high [S~II]/H${\alpha}$ and
[O~III]/H${\beta}$ ratios, \cite{Peimbert88} inferred values of $n_e$
= 1510${\pm}$230 cm$^{-3}$, {\bf $T_e {\sim}80000{\pm}15000$ K}, $v_s
< 160$ km/sec, and an age of ${\sim}$22 kyr.  Based on applied shock
models, \cite{Lozinskaya98} inferred $n_e {\sim}1-10$ cm$^{-3}$, $v_s
{\sim}150-250$ km s$^{-1}$, and an age of ${\sim}3-6$ kyr.  As a
result, they suggested that a combination of slow and fast shocks, as
introduced by \cite{Vancura92} for observations of the LMC SNR N49,
explained the different inferred ages.

\subsection{{\it Chandra} Image}

\begin{figure}
  \caption{{\it Chandra} image of the SNR (upper left), with X-ray
    contours (upper right), (lower left) SNR contours overlaid on an
    H${\alpha}$ image, and (lower right) SNR contours overlaid on a
    merged {\it Swift} UVOT (UVW1) image.  The {\it Chandra} contours
    are drawn at 3.0, 6.0, 9.0, 12.5, and 25.0 counts per pixel.  The
    vertical scale bar is 10 arc seconds in length.  The UVOT image
    demonstrates the number of massive stars surrounding the
    SNR. \label{halpha_snr}}
  \scalebox{0.5}{\includegraphics{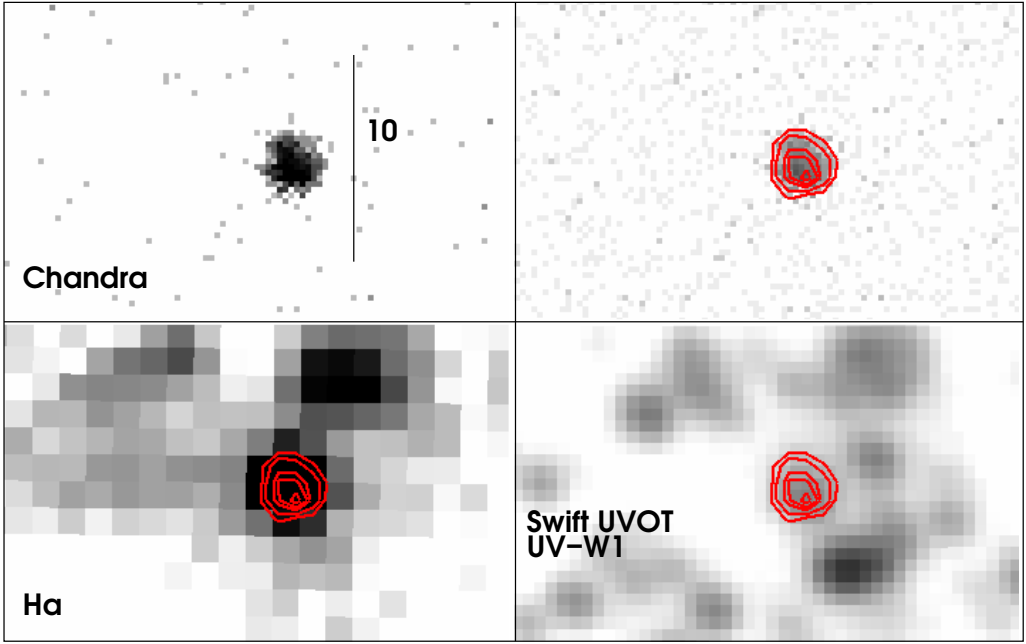}}
\end{figure}

Figure~\ref{halpha_snr} shows a tiling of the {\it Chandra} image, the
{\it Chandra} image with contours, an H${\alpha}$ image, and a merged
{\it Swift} UVOT NUV image (filter UVW1) of the immediate surroundings
of S8.  From the UV image, it is clear that S8 exists in a region of
UV-bright, hence massive, stars.  S8 is resolved in the {\it Chandra}
image: the diameter is ${\sim}5.5''$ which corresponds to
${\approx}$19 pc at the distance to IC~1613.  \cite{Rosado2001}
  infer a diameter of 24 pc for S8 from H${\alpha}$ emission -- that
  the H${\alpha}$ diameter is larger could be attributed to cooling
  gas and snow-plowed matter.  Regardless, either diameter places S8
on the small side of the SNRs in M33 \citep{Long2010} or the LMC
\citep{Ou2018}, for example.

Visually, the X-ray image of S8 resembles a nearly complete circle
with slightly enhanced emission to the south.  There is a hint of a
single high, approximately centered pixel in both images, possibly a
point source (position: 01:05:02.44, +02:08:41.9).  Alternatively,
that location could be a spot of significant interaction in the
ejecta.  A stronger conclusion is not possible at present because of
the distance to S8 and {\it Chandra}'s resolution.  That could change
with the detection of a compact object, for example.  The {\it
  Chandra} image then suggests that S8 could be a plerion.  We return
to that identification in \S\ref{Disc}.

\cite{Lozinskaya98} described S8 as resembling the LMC SNR N~49.  In
comparison with SNR catalogs (e.g., \citealt{Williams1999}), however,
we think a more appropriate comparison lies with N~63A (e.g.,
\cite{Warren2003}): in N~49, there is a strong interaction in one
direction (southeast) with a significant decrease in emission in the
opposite direction.  For S8, there is enhanced emission to the south,
but the drop in the opposite direction is less steep -- a description
that mimics N~63A.  \cite{Rosado2001} also make the comparison
  between S8 and N~63A.

\begin{figure}
\caption{Hardness image of the SNR where Hardness = (H - S ) / ( H + S )
and the bands H and S are defined as H = 0.75 - 2 keV and S = 0.4 - 0.75 keV.
The vertical scale bar is 7 arc seconds in length.}
\label{snr_Hard}
\scalebox{0.6}{\includegraphics{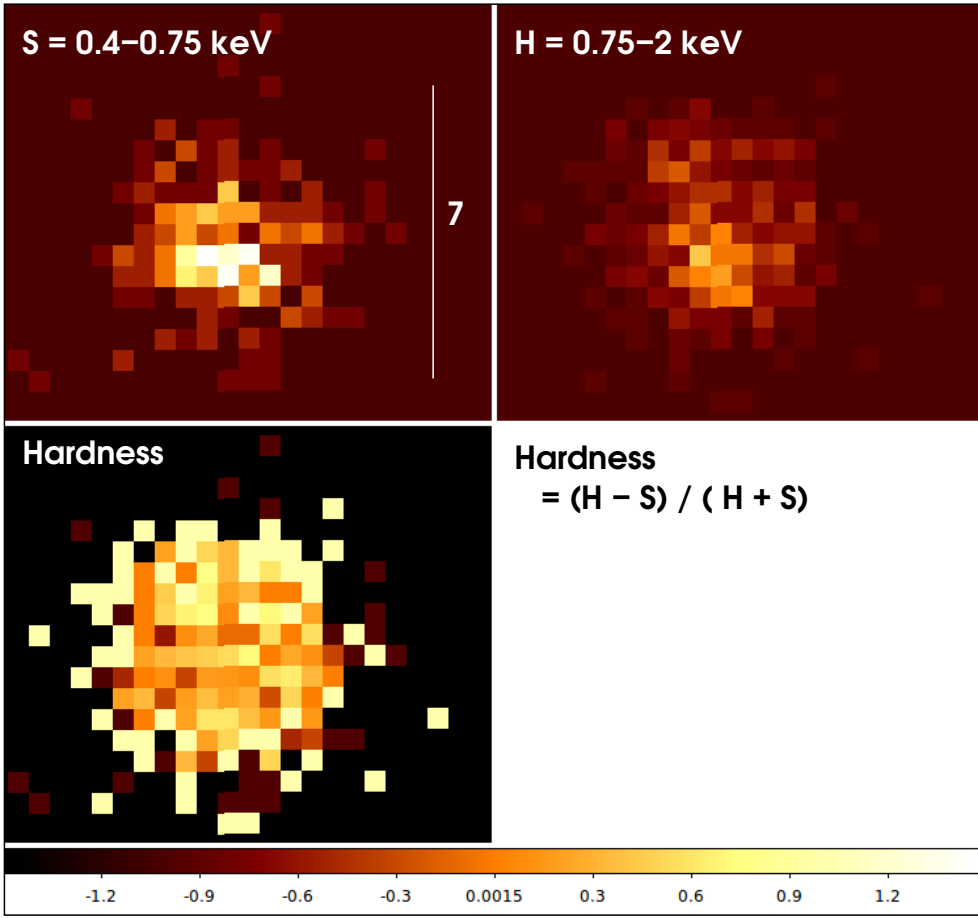}}
\end{figure}

Figure~\ref{snr_Hard} shows a hardness image of S8 where the hardness
is defined as $HR = {{H - S}\over{H + S}}$ and where the bands are
defined as S = 0.4 - 0.75 keV and H = 0.75 - 2 keV.  These energy
  ranges were chosen to maximize the available spectral information
  based on the observed spectrum (\S\ref{Xrays}): they separate low-Z
  lines (N, O) from higher-Z lines (Ne, Si, Fe L) typically found in
  SNRs.   The hardness image runs from -1 to +1 and the color coding
of the image is so defined.  Note in the image that there is a hard,
nearly circular edge to the SNR while the interior is considerably
softer.  The bright central spot does not stand out for being very
hard nor very soft.  The color coding at that location corresponds to
${\approx}$0.3 in $HR$, suggesting that if there is a point source
present, it is not a typical hard pulsar like the Crab.

\begin{figure}
\caption{Radial profiles as cuts along the SE, SW, NW, and NE radii.
 Each cut represents a 45-degree-wide slice centered on the
 above-listed directions with eight radial bins.  The horizontal scale
 is in {\it Chandra} pixels.  The profiles are slightly offset horizontally
 from each other for visibility.  The squares indicate the {\it Chandra}
 PSF \citep{Jerius2000} and demonstrate that {\it Chandra} easily resolves S8.}
\label{snr_profile}
\scalebox{0.5}{\rotatebox{-90}{\includegraphics{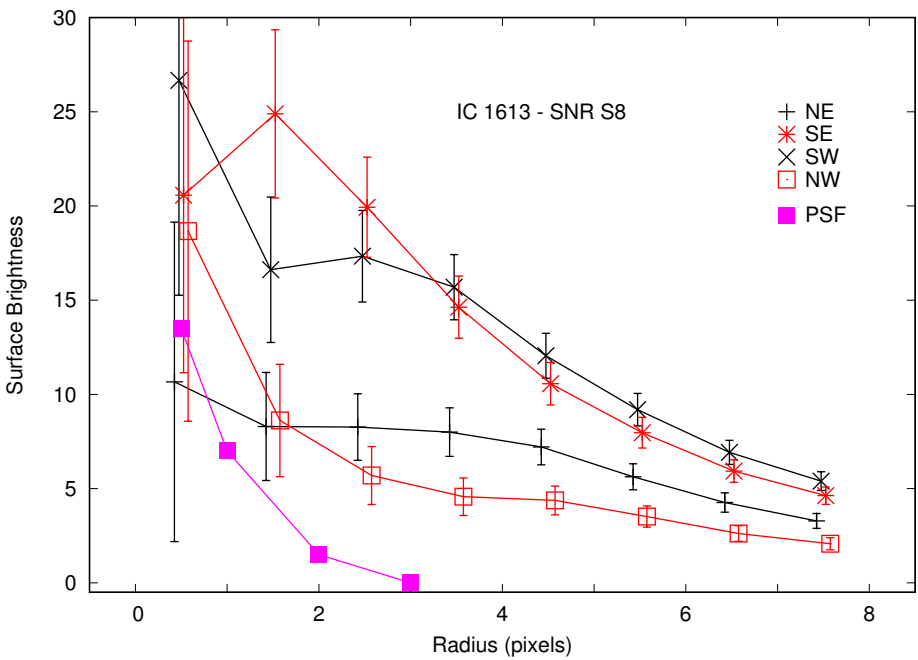}}}
\end{figure}

Figure~\ref{snr_profile} shows radial profiles of the {\it Chandra}
image, quantifying the shape of the shell.  Given the circular
  nature of S8, the radial profiles were extracted from a center
  determined by the mean circular outer edge.  The surface
brightnesses along the SE, NE, SW, and NW radii were then plotted.
The choice of direction for the radial cuts was made to test the
elliptical shape reported by \cite{Lozinskaya98} for the HRI and
H${\alpha}$ images.  There are differences in the profiles, but most
of the differences are confined to the interior.

That the profiles demonstrate approximately equal emission throughout
the remnant argues for a plerionic or young composite interpretation
of the SNR.  \cite{Rosado2001} argued against that interpretation,
based on the available {\it ROSAT} data as well as optical and radio
images.  They suggested that S8 represented only part of a shell of an
older remnant, with much of the SNR hidden by dust.  Given {\it
  Chandra}'s spatial resolution and the observation presented in this
paper, we do not agree with \cite{Rosado2001}.

\subsection{Radio and H${\alpha}$ Images}

\begin{figure} 
\caption{Comparison of an H${\alpha}$ position-velocity cube
(grayscale + red contours) from \cite{Lozinskaya03} of S8 with the
{\it Chandra} X-ray data (cyan contours).  The numbers in each
sub-frame represent the velocity (km/sec) at the `position' within the
data cube as noted next.  A subset of the data cube are presented,
representing the (left to right, top to bottom) 16\% (16) and 84\%
(84) of the flux of the blue peak, the blue peak (B), the red peak
(R), and the 84\% (84) and 16\% (16) flux of the red peak.  The {\it
Chandra} contours occur at fluxes of 3, 6, 9, 12.5, and 25 counts per
pixel (as in Figure~\ref{halpha_snr}).  The red contours in both 16\%
images occur at 100, 150, 200, 250, 300, and 350 (units of $10^{-15}$
erg s$^{-1}$ cm$^{-2}$ {\AA}$^{-1}$).  In both 84\% images, the
contours occur at 300, 500, 750, 1000, 1250, 1500, and 1650 in the
same units. For both peaks, the `1650' contour is replaced with
contours at 1750 and 2000.  Details of the H${\alpha}$ image are
described in \cite{Lozinskaya08}.}
\label{Half}
\scalebox{0.6}{\includegraphics{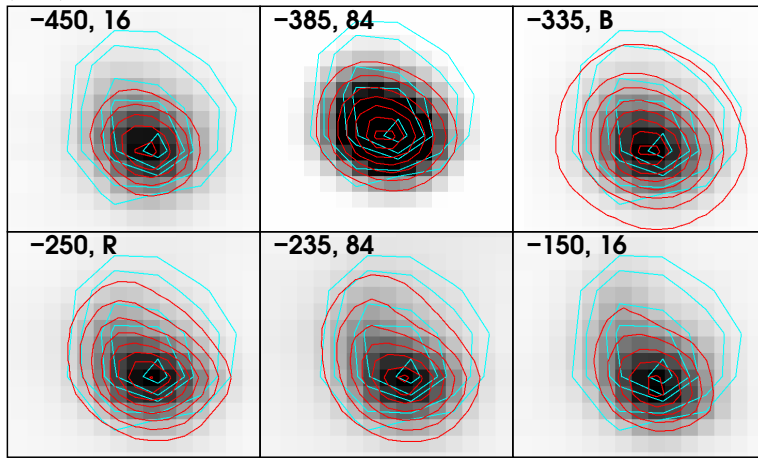}}
\end{figure}

S8 has been observed in the radio and H${\alpha}$ bands and it is
resolved in both bands
(\citealt{Lozinskaya03,Lozinskaya08,Rosado2001}).  Figure~\ref{Half}
shows contours from the {\it Chandra} image overlaid on an H${\alpha}$
data cube described by \cite{Lozinskaya03}.  While the peak of the
radio emission consistently lies east of the X-ray peak, it does not
appear to be spatially distinct or resolved. The H${\alpha}$ data cube
of \cite{Rosado2001} is very similar.  Overall, the X-ray and
H${\alpha}$ contours are spatially coincident and do not demonstrate
significant distortion from a circular form.  It is possible that a
higher {\it spatial} resolution version of Figure~\ref{Half} could
determine the location of a compact object.  Certainly that the peaks
are approximately in the same location warrants a sensitive,
high-resolution observation at that location.

A comparison of the {\it Chandra} and radio data yield an image
similar to Figure~\ref{Half}, which is not included here.  S8 is much
less luminous than other radio SNRs in our Galaxy and nearby irregular
galaxies: the radio luminosity of S8 is 5\% and 17\% of the radio
luminosity of Cas A and the Crab respectively, and 15\% of the radio
luminosity of LMC N 49, the brightest SNR in the Large Magellanic
Cloud (\cite{Becker91}, \cite{Reed95}, \cite{Feast91}).

The lower radio luminosity of S8 is probably due to the SNR shock
expanding into a low-density ISM as shown by the H~I measurements of
\cite{Lozinskaya08}.  The SNR is positioned on the edges of a series
of H~I arcs as noted in their data.

\section{The X-ray Spectrum}\label{Xrays}

\begin{figure}
  \caption{Extracted spectrum of the entire SNR showing the best-fit {\tt vapec} model
    from Table~\ref{SpecFit}.}
\label{spec_snr}
\scalebox{0.5}{\rotatebox{-90}{\includegraphics{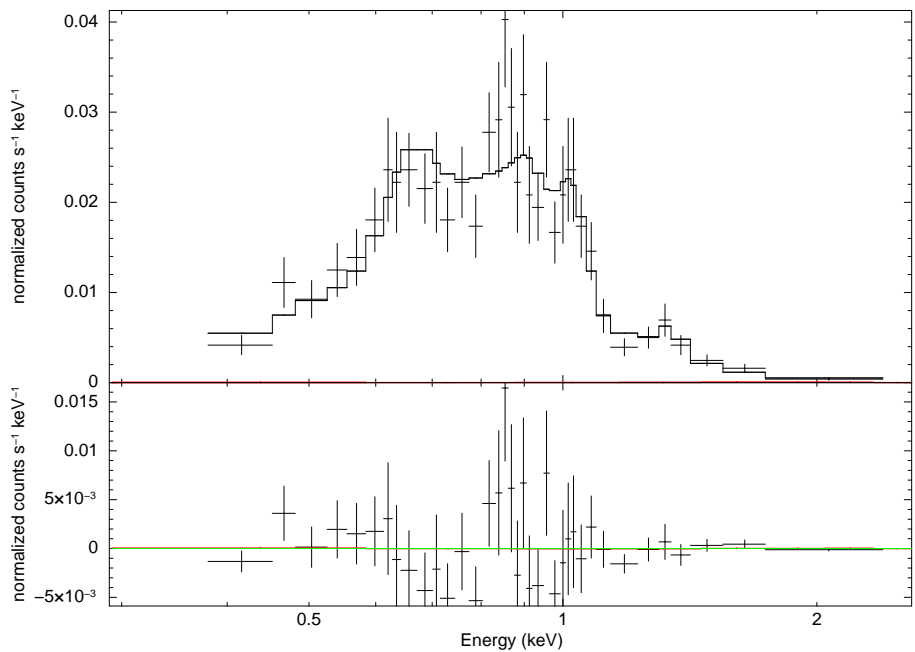}}}
\end{figure}

The spectrum shows considerable emission in the 0.7-1.0 keV band,
likely un-resolved emission lines of Fe or Ne
(Figure~\ref{spec_snr}(a)).  There are zero events below $\sim$0.4 keV
or above ${\sim}$3-4 keV.

We carried out spectral fits using {\it XSPEC} \citep{Arnaud96} using
the `cstat' statistic.  Fitted parameter values are listed in
Table~\ref{SpecFit}.

With an absorption component included, we tried several different
models including powerlaw, thermal {\it bremsstrahlung}, optically
thin thermal gas models ({\tt apec, vapec}, where the `v' indicates
individually-variable abundances) and Sedov shock models ({\tt Sedov,
  vSedov}; \citealt{Borkowski2001}).  The absorption component used
the \cite{Wilms2000} abundance values.

The brems and powerlaw fits provided poor descriptions of the spectrum
and are not included in Table~\ref{SpecFit}.  The powerlaw index
was steep (${\sim}7.7{\pm}0.9$) and that steepness establishes the
spectrum as a thermal one, as expected from an SNR.

The best-fit models were the variable-abundance optically thin hot gas
model (\citealt{Foster2010,Smith2001}) and the Sedov shock model
\citep{Borkowski2001}.  We tested individually-varying abundances in
the {\tt vsedov} model, but found that all of the abundance
differences were consistent with solar.  Consequently, we do not
include this model in Table~\ref{SpecFit}.

The remaining two models ({\tt vapec}, {\tt Sedov}) describe the
spectrum equally well, but the fitted parameters differ.  One may
criticise the use of the `variable' variant of the {\tt vapec} model
given the number of counts in the spectrum.  We argue however that the
determination of possible non-solar abundances outweighs the low
signal-to-noise.  We do {\it not} ignore the low signal-to-noise as we
fit or fix the abundance of each element sequentially rather than
simultaneously.  

Furthermore, the non-variant {\tt apec} model delivers a visibly poor
fit with clear residuals concentrated near the O lines at 0.6 keV and
the Fe/Ne lines at 1 keV.  We infer from this that, to obtain a good
fit to the spectrum, we need at least one additional parameter beyond
varying the temperature and column density - whether varying an
abundance ({\tt vapec}) or an ionization time ({\tt Sedov}).  This
then also explains why we do not see significant abundance variations
with the varying-abundance {\tt vSedov} model: much of the power of an
added parameter is taken up by the ionization time in the {\tt Sedov}
variants.  Consequently, additionally varying abundances does not lead
to a result significantly different from solar.

\begin{figure}
\caption{(a) Confidence contours for the {\tt vapec} fit (shown in
Figure \ref{spec_snr}) for SNR parameters kT and N$_{\rm H}$.  The known
Galactic column is ${\approx}$0.02 on this plot.\label{SNR_cont}}
\scalebox{0.5}{\rotatebox{-90}{\includegraphics{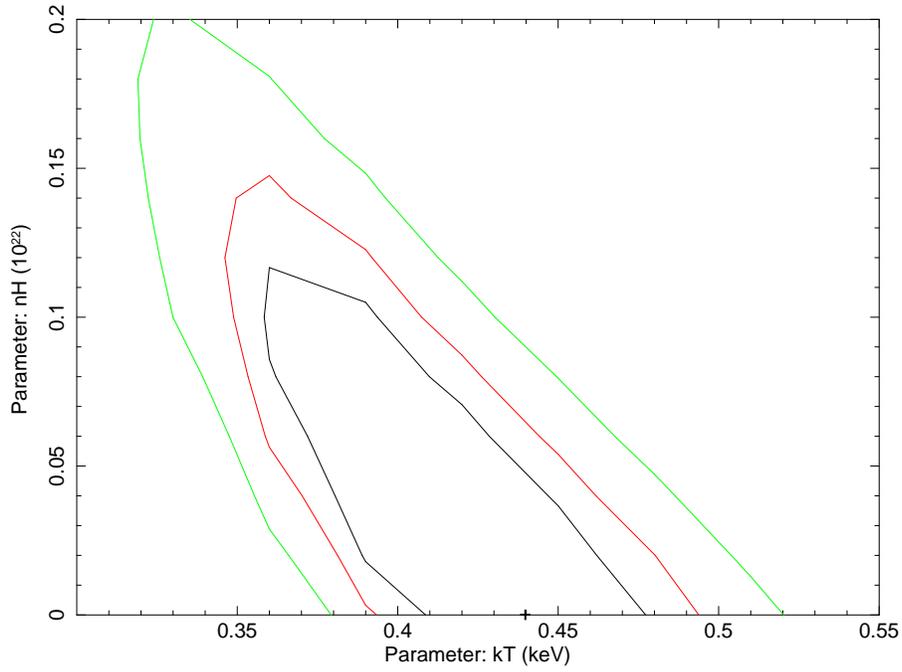}}}
\end{figure}

The remainder of the spectral discussion will focus on just these two
models.  Figure~\ref{SNR_cont}(a) shows the contours on N$_H$ and kT
for the {\tt vapec} fit and are representative of similar plots for
the {\tt Sedov} model.

The `fitted' column density is an upper limit: the best values are
${\lesssim}0.9{\times}10^{21}$ cm$^{-2}$ ({\tt Sedov}) and
${\lesssim}1.5{\times}10^{21}$ cm$^{-2}$ ({\tt vApec}) essentially
because of the lack of events below $\sim$0.4 keV.  If we assume the
column density lies immediately below those limits, then using the
$N_H-E_{(B-V)}$ relation of \cite{PS95} leads to an estimate of
$E_{(B-V)} {\sim}0.17-0.28$.  Given that IC~1613 is a low-metallicity
galaxy, we can also argue for the adoption of the $N_H-E_{(B-V)}$ of
\cite{Fitzpatrick1985} (for the SMC) which leads to $E_{(B-V)} {\sim}
0.11-0.17$.  Those values represent an enhancement in the column by a
factor of $\approx$3-8 above the known Galactic column toward
IC~1613.

\cite{Rosado2001} infer that the SNR could be partially hidden by
dust, so an enhanced column density should be expected.  No values are
given in their paper.  If, however, we crudely assume a spheroid or
cuboid shape for IC 1613 and that S8 lies roughly on the mid-plane,
then the column length within IC 1613 is ${\approx}5$ kpc based on the
tidal radius of \cite{Battinelli2007}.  With a hydrogen number density
of 1 cm$^{-3}$ across 1 kpc yielding a column of
${\approx}3{\times}10^{21}$ {\rm cm}$^{-2}$, the numbers work: either
a number density of a few per cm$^{-3}$ over a shorter column or a
lower value (${\approx}0.1 {\rm cm}^{-3}$) over a longer column.
Given the appearance of the region surrounding S8 as well as the lower
H I density in the core of IC 1613 compared to the NE complex
\citep{Lake1989,Berger2018}, the first case is the more likely, with
dust in the region of the NE complex.

The fitted temperatures are ${\sim}$0.43 keV for the {\tt vapec} model
and ${\sim}$0.98 keV for the {\tt Sedov} model.  The uncertainty on
the {\tt vapec} model is ${\approx}$15\% vs a significantly larger
uncertainty of ${\approx}$30-80\% for the {\tt Sedov} model.  Lacking
a basis to choose one over the other, we will explore the implications
of temperatures in the 0.4-1 keV range.

We also checked for altered abundances.  The simple {\tt Sedov} model
provides an upper limit that is consistent with solar, while the
variable {\tt vapec} model exhibits two non-solar abundances.
Figure~\ref{FeOcont}(b) shows the contours for the {\tt vapec} model
for the two elements with significant non-solar abundances: O and Fe.
The oxygen value is ${\approx}$0.5 with an uncertainty of ${\sim}$60\%
while the Fe value is ${\approx}$0.15 with an uncertainty of
${\sim}$40\%.  

The {\tt Sedov} model has an extra parameter, the ionization time, the
product of the SNR's age and its post-shock electron number
density. Both the simple and the variable-abundance versions returned
fitted values of the ionization time ${\sim}8.5-9.5{\times}10^{11} {\rm s cm}^{-3}$.
This indicates that the SNR is close to collisional ionization
equilibrium (CIE), validated by the acceptable fit with the {\tt
vapec} model, which itself assumes CIE.  That value fits within the
picture described in \cite{Smith2010}.

The integrated flux of the SNR in the 0.3-8 keV band for the best-fit
models is ${\approx}5.2-5.3{\times}10^{-14}$ erg s$^{-1}$ cm$^{-2}$;
the unfolded flux in that band is $5.4-5.6{\times}10^{-14}$ erg
s$^{-1}$ cm$^{-2}$ corresponding to an unfolded luminosity of
${\approx}5.6{\times}10^{36}$ erg s$^{-1}$.  Our integrated L$_X$ is
25\% higher than the estimate in \cite{Lozinskaya98}.  In that paper,
the flux was estimated in the 0.1-2.4 keV for the {\it ROSAT} HRI, an
instrument without spectral resolution, using several basic continuum
models (e.g., {\it bremsstrahlung}, Raymond-Smith, power law).  As we
have shown, there is little flux below ${\sim}$0.4 keV nor above
${\sim}2.5$ keV, so the two estimates should be considered to be
identical.

\section{Discussion}\label{Disc}

Shock physics allows us to take the numbers we have measured and turn
them into estimates of the explosion energy and age of the SNR.  We
use expressions from \cite{Lequeux2005} for the shock radius $r_s$ and
the shock temperature $T_s$:

 $$r_s = 0.26 \left({{n_H}\over{cm^{-3}}}\right)^{-{{1}\over{5}}} \left({{t}\over{yr}}\right)^{{{2}\over{5}}} \left({{E}\over{4{\times}10^{50} ergs}}\right)^{{{1}\over{5}}}~pc$$

 $$T_s = 1.5{\times}10^{11} \left({{n_H}\over{cm^{-3}}}\right)^{-{{2}\over{5}}} \left({{t}\over{yr}}\right)^{-{{6}\over{5}}} \left({{E}\over{4{\times}10^{50} ergs}}\right)^{{{1}\over{5}}}~K$$

{\noindent}where $n_H =$ number density; $t =$ age; and $E =$
explosion energy.  We have measures for $r_s$ and $T_s$; we need an
estimate for the number density to proceed.  The model normalization
provides one estimate for $n_H$; another estimate comes from the
optical [S~II]/H${\alpha}$ ratio.

The model normalizations are scaled identically with the normalization
defined equal to 
   $${{10^{-14}}\over{4 {\pi} [D_A(1 + z)]^2}} \int n_e n_H dV.$$

{\noindent}where $D_A =$ angular diameter distance; $n_e = 1.2 n_H =$
number densities of electrons and protons, respectively; $V =$
emission volume; and $z =$ redshift = 0 for IC~1613.  Inserting
numbers leads to an estimate of $n_H {\sim} 1.6{\pm}0.2$ cm$^{-3}$ for
the {\tt vapec} and {\tt Sedov} models.

With the number density estimates, we are then left with two equations
with two unknowns: the age $t$ and the explosion energy $E$.
Inserting all numbers leads to

$$t {\sim} 3380-5650~{\rm yr}$$ and $$E {\sim}3.5-9.9{\times}10^{51}~{\rm ergs}$$

{\noindent}for the age and explosion energy, respectively, and where
the range results from the range in temperature (0.4-0.98 keV).  Age
and explosion energy trade off here, so if we use the estimate of the
optical number density, the age range shrinks to 1440-2410
yr and the explosion energy becomes ridiculously high, with a {\it
lower} limit of 10$^{54}$ ergs.  This is clearly non-sensical.  We
therefore conclude that the number density in the X-ray-emitting
region is low, with $n_H {\sim}1.5-2$ cm$^{-3}$, and confirming the
results of \cite{Lozinskaya98}.

With those numbers in hand, the shock velocity is then ${\sim}$660
km/sec for the {\tt vapec} model or ${\sim}$1100 km/sec if we use the
{\tt Sedov} model numbers.  Based on the analysis of
\cite{Lozinskaya98}, the lower value is favored.

In Table~\ref{CompTab}, several basic properties of S8 are compared
with other Local Group SNRs.  For similar ages of 3-6 kyr, the number
densities generally lie in the ${\approx}$0.1-1 cm$^{-3}$ range.  The
luminosity of S8 is the least luminous of the SNRs in the table,
likely reflecting the lower ISM density in a dwarf galaxy.

We can assign an upper limit on the presence of a pulsar at the center
of S8.  We adopt two different types of central object: a Crab-like
pulsar and a Cas A-like compact object.  Assuming that we may treat a
compact object as filling a single {\it Chandra} pixel and because we
do not know the emission properties in the center of the SNR, the
detected flux in the center constrains our upper limit.  The counts in
those two pixels are 35${\pm}$7.0, for a count rate of
${\sim}7.1{\pm}1.4{\times}10^{-4}$ cts s$^{-1}$ and 53${\pm}$8.3 for a
count rate of ${\sim}1.1{\pm}0.2{\times}10^{-3}$ cts s$^{-1}$.  Both
values are well-above the interpolated background counts at the
pixels' locations and above the limiting sensitivity values for a
point source.

Adopting a soft spectrum appropriate for a Cas A-like object leads to
an upper flux limit of ${\sim}5.7{\times}10^{-15}$ erg s$^{-1}$
cm$^{-2}$ or a 0.3-5 keV luminosity of ${\sim}3.6{\times}10^{35}$ ergs
s$^{-1}$.  A Cas A-like object has a luminosity in a similar band of
${\sim}{\rm few}{\times}10^{33}$ erg s$^{-1}$ \citep{Chakrabarty2001}.
On the basis of this estimate, we could not detect a Cas A-like
object, failing by ${\approx}$2 orders of magnitude.  Using those same
numbers, however, we would easily detect a Crab-like pulsar with a
typical luminosity 10-100 times above our sensitivity value.  If we
estimate the flux from a harder spectrum more typical of a Crab-like
pulsar, then our limiting luminosity rises slightly to
${\sim}9{\times}10^{35}$ erg s$^{-1}$, but still well below the actual
luminosities of Crab-like pulsars, so, again, easily detected.

We note that S8 is detected in the X-ray, optical, and radio
bands and behaves rather similarly in all of them.  This behavior is
counter to the conclusion of \cite{Pannuti2007} in which detections of
SNRs in nearby galaxies in {\it all} three bands form a very sparse
set, in contrast to the number of SNRs detected in any one of those
bands.  Without additional data, the explanation could always be
attributed to insufficient sensitivity of the detectors.  However, the
observations presented here of S8 at least partially demonstrate that
the necessary sensitivity is present, again raising the puzzle 
described by \cite{Pannuti2007}.

Given the X-ray, optical, and radio detections, what type of SNR is
S8?  The circular, filled X-ray image argues for a plerion or young
composite SNR.  The spectral indices of the two radio observations,
while consistent with each other, are {\it not} consistent with a
plerion which have typical values in the 0.05-0.30 range (for 3C 58
\citep{Bietenholz2001} and the Crab \citep{Bietenholz91,Baars1977})
compared to S8's ${\approx}-0.6~{\rm to}~ -0.7$ index value.  We then
infer that S8 is a young composite SNR.

Do irregular or dwarf galaxies affect the evolving SNRs in a manner
different from spirals?  \cite{Bozzetto2017} displayed Venn diagrams
for the irregular galaxies SMC, LMC, NGC 4449, NGC 3077, NGC 4214, and
NGC 5204.  All except the LMC and the SMC showed minimal overlap in
radio, optical, and X-ray bands.  The answer to the question would
then appear to be `no,' which means the original puzzle remains.
Clearly more work and higher sensitivity surveys are needed to
understand the behaviors of SNRs across the EM spectrum.

\cite{Ou2018} examined LMC SNRs on the basis of their size and X-ray
luminosity.  They showed that small SNRs all had Type Ia progenitors
while medium-sized SNRs were dominated by Type II progenitors because
the massive stars had evacuated the immediate vicinity of space.  On
the basis of its size and L$_{\rm X}$, S8 falls in a group of five
remnants (N132D, N63A, 0540-69.3, N49, and DEM L71) only one of which
had a Type Ia progenitor.  On the basis of the its appearance, and its
size and luminosity, the position of S8 in the \cite{Ou2018} plot
suggests it had a massive progenitor.

\section{Summary}

We have described the {\it Chandra} observation of S8, the only known
SNR in the irregular Local Group galaxy IC~1613.  S8 is visible in the
X-ray, H${\alpha}$, and radio bands.  The SNR is resolved with {\it
  Chandra} into a roughly circular structure with a diameter of
${\sim}$19 pc and an estimated age of ${\sim}$3400-5600 years.  It
exhibits enhanced X-ray emission south of the center of the SNR as
defined by the roughly circular edge, but the enhancement is
relatively soft.  There may be a source at or near the center based on
the enhanced emission in the region.  Folding together the radio
spectral index, the circular outer boundary in the X-ray image, the
apparently filled appearance in the center, we suggest S8 is a young
composite SNR.

\vspace{0.5in}
\facility{CXO(ACIS)}

\acknowledgements

We thank the referee for comments that improved this paper.  This
research has made use of NASA's Astrophysics Data System as well as
the NASA/IPAC Extragalactic Database (NED) which is operated by the
Jet Propulsion Laboratory, California Institute of Technology, under
contract with the National Aeronautics and Space Administration. This
research has also made use of data obtained from the High Energy
Astrophysics Science Archive Research Center (HEASARC), provided by
NASA's Goddard Space Flight Center. This work was supported by Chandra
Grant GO3-4104Z.

\clearpage

\begin{table}
\footnotesize
\begin{center}
\caption{Spectral Fits for the SNR$^a$\label{SpecFit}}
\begin{tabular}{lrrrrrrrrrr}
        &     &       &              &               &                     &            &               &             & Flux$^c$  & Unfolded Flux$^c$ \\
        &     &       &              & Primary       & $N_{\rm{H}}$        & Other      &               & Model       & 0.5-2 keV & 0.5-2 keV \\
 Model  & DoF & cstat & $\chi^2$/DoF & Parameter$^b$ & 10$^{22}$ cm$^{-2}$ & Parameter$^b$  & Abundance & Norm$^a$    & 2-8 keV   & 2-8 keV   \\ \hline
\\
Apec    & 39 &  79.4 & 2.06      & T: 0.59(4)      & $<$0.6          & ${\cdots}$       & all: $<$1.10 & 1.4(6)[-5] & ${\cdots}$ & ${\cdots}$ \\
v-Apec  & 37 &  45.3 & 1.05      & T: 0.43(6)      & $<$0.15         & ${\cdots}$       & O: 0.46(26)  & 5.(9)[-5]  & 5.2[-14] & 5.6[-14] \\
        &    &       &           &                 &                 &                  & Fe: 0.15(6)  &            & 1.2[-14] & 1.2[-14] \\ 
Sedov   & 38 &  35.3 & 0.81      & T: 0.98$^{+0.31}_{-0.61}$ & $<$0.09 & T$_b$: $<$1.4  & all: $<$1.05 & 5.4[-5]    & 5.3[-14] & 5.5[-14] \\
        &    &       &           &                 &                 & ${\tau}$: 8.5$^{+3.7}_{-2.8}$[11] &          & 1.3[-14] & 1.3[-14] \\
\hline
\end{tabular}

$^a$Numbers in () = symmetric uncertainty in last digit(s); numbers in [] =
$10^{XX}$; 

$^b$Units of T and T$_b$ = keV; units of ${\tau}$ = s cm$^{-3}$.

$^c$Fluxes for best-fit models only; all fluxes in ergs s$^{-1}$ cm$^{-2}$.

\end{center}
\end{table}

\normalsize

\begin{table}
\begin{center}
\caption{Comparison of SNR S8 with Other Local Group SNRs$^a$}
\label{CompTab}
\begin{tabular}{lrrrrrrr}  
\hline
 SNR           & T$_s$   & N$_H$               & L$_x$         & D$_s$        & n$_H$  & Age   & E$_0$ \\ \hline
 S8$^b$        & 0.43    & $<$1.5              &  3.5         & 19          & 1.7    & 3.8    & 2 \\
               & 0.98    & $<$0.9              & $\cdots$     & ${\cdots}$  & 1.8    & 6.5    & 7 \\
 NGC 6822$^c$  & 2.8$^{+6.1}_{-2.0}$ & $3.0^{+1.9}_{-0.7}$ & 8.1 & 24 & 0.03$^{+0.06}_{-0.02}$ & 3.02$^{+2.74}_{-1.32}$ & 0.3AD \\ 
 M33 SNR21$^d$ & 0.46(2) & 0.5+$<$0.3$^d$       & 15           & 10          & 1.7    & 6.7     & 1.8 \\
 LMC N63A$^e$  & 0.6     & 1.7${\pm}$0.1       & $\cdots$     & 16.4        &  5     & few$^e$  & $\cdots$ \\
\hline
\end{tabular}
\end{center}

$^a$Units: T in keV; N$_H$ = absorption column in units of $10^{21}$
  cm$^{-2}$; L$_x$ units = $10^{36}$ erg s$^{-1}$ in the 0.5-8 keV
  band; D$_s$ = SNR diameter in pc; $n_H$ = number density in
  cm$^{-3}$; Age in kyrs; E$_0$ = explosion energy in units of
  $10^{51}$ ergs; AD = ADopted value.

$^b$This paper.

$^c$From \cite{Kong2004}.

$^d$From \cite{Gaetz2007} using the entire SNR field 4 dataset for the {\tt sedov} model. The
  N$_H$ values represent the Galactic column + the local column.

$^e$From \cite{Chu1999} and \cite{Warren2003}.
\end{table}


\setcounter{figure}{5}
\begin{figure}
\caption{(b) Confidence contours for the SNR {\tt vapec} abundance
parameters for Fe and O.    \label{FeOcont}}
\scalebox{0.5}{\rotatebox{-90}{\includegraphics{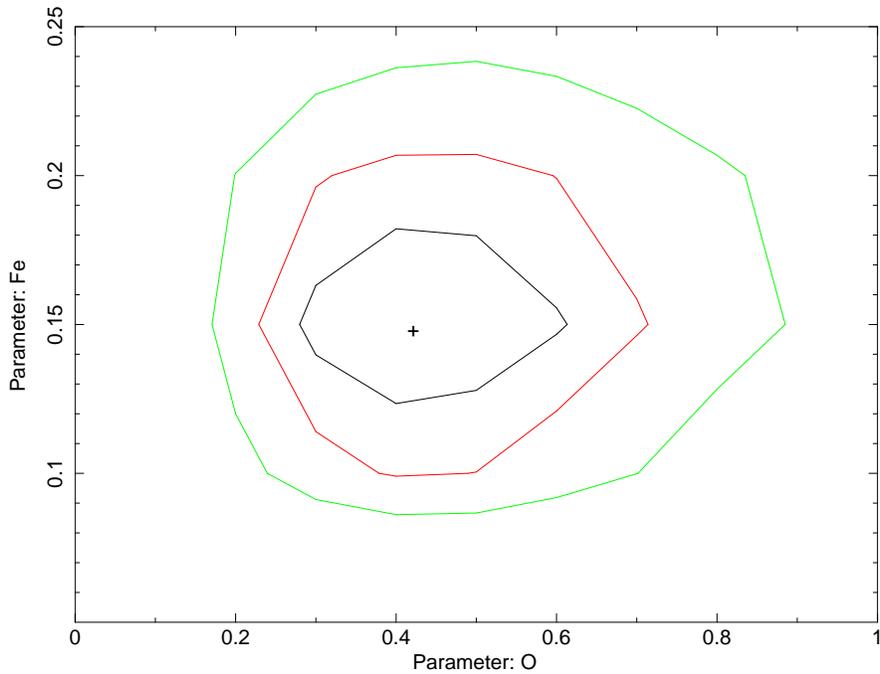}}}
\end{figure}

\end{document}